\documentclass[12pt]{article}

\usepackage[body={17.5cm, 23cm},right=2cm]{geometry}

\usepackage{booktabs}
\usepackage{caption}

\usepackage{color}
\usepackage{graphicx}
\usepackage{epsf}
\usepackage{graphicx,epsfig}
\usepackage{multirow}

\usepackage{amsmath, amsthm, amssymb}

\usepackage{booktabs}

\usepackage{epsfig}
\usepackage{cite}
\usepackage{color,colordvi}

\newcommand{\be}{\begin{eqnarray}}
\newcommand{\ee}{\end{eqnarray}}

\newcounter{hran}

\makeatletter
\renewcommand\section{\@startsection {section}{1}{\z@}%
                               {-3.5ex \@plus -1ex \@minus -.2ex}%
                               {2.3ex \@plus.2ex}%
                               {\normalfont\large\bfseries}}
\makeatother

\numberwithin{equation}{section}
\begin{document}
\vspace{5mm}
\vspace{0.5cm}

\vspace{5cm}

\begin{center}

\def\thefootnote{\fnsymbol{footnote}}

{\large \bf 
  Emergent Potentials in Consistent Higher Derivative $N=1$ Supergravity
  %
}
\\[1.5cm]
{\large  Fotis Farakos and Alex Kehagias }
\\[0.5cm]

\vspace{.3cm}
{\normalsize {\it  Physics Division, National Technical University of Athens, \\15780 Zografou Campus, Athens, Greece}}\\

\vspace{.3cm}
{\normalsize { E-mail: fotisf@mail.ntua.gr,  kehagias@central.ntua.gr }}


\end{center}

\vspace{3cm}

\hrule \vspace{0.3cm}
\small  \noindent \textbf{Abstract} 
\noindent

By employing consistent supersymmetric higher derivative terms, 
we show that the supersymmetric theories may have a sector where the scalar potential does no longer 
have the conventional form.
 The theories under consideration contain consistent
higher-derivative terms which do not give rise to instabilities and ghost
states. The chiral
auxiliaries are still not propagating and can be integrated out. Their
elimination gives rise to
emerging potentials even when there is not a superpotential to start with.
This novel feature of higher derivative supersymmetric chiral models is also
extended to vector multiplets both in global and local supersymmetry. In
particular, in supergravity,  the emerging potentials
give rise always to  a de Sitter vacuum signaling supersymmetry breaking.

\vspace{0.5cm}  \hrule
\vskip 2.5in

\noindent
{\it Proceedings of the Corfu Summer Institute 2012 \\
 		 September 8-27, 2012\\
 		 Corfu, Greece}

\def\thefootnote{\arabic{footnote}}
\setcounter{footnote}{0}


\newpage



\baselineskip= 19pt

\section{Introduction: Scalar Potentials in $N=1$ Superspace}

Supersymmetry is an extension of the Poincare spacetime symmetry with the
inclusion of fermionic generators.
It has various remarkable properties concerning phenomenological and
theoretical aspects of particle physics.
In particular, supersymmetry is one of the most appealing candidates for
new physics. It has not been observed so far
and thus, it should be broken at some high energy scale if it is realised
at all.
The central role on how supersymmetry is broken is played by the scalar
potential of the supersymmetry breaking sector.
Scalar potentials in supersymmetry and supergravity have been extensively
studied for theories with up to two derivatives.
Even though it is known that introducing higher derivatives will spoil the
form of the scalar potential,
the self-consistency of the theory protects it from unconventional
non-supersymmetric vacua \cite{Cecotti:1986jy}.
Our task here is to discuss how scalar potentials are modified when higher
derivatives are introduced. However,
the higher derivatives we are interested in, are those which do not
introduce instabilities and/or ghost states.
This is a known
drawback of such kind of interactions, 
connected with the so-called Ostrogradski \cite{ostro} instability in classical physics.
We will see that such ``safe'' higher derivatives may consistently be
introduced in supergravity  and we will determine
the form of the potential for the scalars of the theory they produce.
We will also see that such potentias are sustained by background fluxes and
have de Sitter vacua indicating that supersymmetry
is broken.

In this work we are discussing the bosonic sector of supersymmetric interactions that belong 
to a specific class of higher derivative theories with the following two properties
\begin{enumerate}
 \item 
they do not introduce ghost states
\item
they introduce a scalar potential without a superpotential or gauging.
\end{enumerate}
These theories involve chiral and vector multiplets.

In $N=1$ superspace there is a number of conventional methods to introduce a scalar potential for a chiral superfield.
The {\it superpotential} is the most widely used, 
in which case one employes a holomorphic function of the chiral superfield 
and after integrating out the auxiliary sector, a scalar potential appears.
More specificaly, the free Wess-Zumino Lagrangian is given by \cite{Wess:1992cp}
\be
\label{cL0}
L_{0} = A \partial^2 \bar{A} 
+ i \partial_{a} \bar{\psi}_{\dot{\alpha}} \bar{\sigma}^{a \dot{\alpha} \alpha } \psi_{\alpha}  
+ F \bar{F}.
\ee
It is straightforward to integrate out the auxiliary field via its equations of motion 
\be
\label{F1}
F=0 
\ee
which for the massless and free theory (\ref{cL0}) vanishes,
leading to
\be
\label{cL1}
L_{0} = A \partial^2 \bar{A} 
+ i \partial_{a} \bar{\psi}_{\dot{\alpha}} \bar{\sigma}^{a \dot{\alpha} \alpha } \psi_{\alpha}.
\ee
A standard mass term contribution is given by employing the following Lagrangian
\be
\nonumber
L_{0} + \frac{m}{2} (L_{m} + h.c.) &=& A \partial^2 \bar{A} 
+ i \partial_{a} \bar{\psi}_{\dot{\alpha}} \bar{\sigma}^{a \dot{\alpha} \alpha } \psi_{\alpha}  
+ F \bar{F} 
\\
\label{T1}
&+& m  F A - \frac{1}{2} m \psi^{\alpha} \psi_{\alpha} 
+ m \bar{F} \bar{A} - \frac{1}{2} m \bar{\psi}_{\dot{\alpha}} \bar{\psi}^{\dot{\alpha}}.
\ee
A naive inspection of (\ref{T1}) would tell us that there is massive fermions, 
but no mass for the scalar fields has appeared.
The equations of motion for the auxiliary field $F$ read
\be
\label{F2}
\bar{F}= - m A
\ee
and eventually, the on-shell form of (\ref{T1}) becomes
\be
\label{T2}
L_{0} + \frac{m}{2} (L_{m} + h.c.)  = A \partial^2 \bar{A} 
+ i \partial_{a} \bar{\psi}_{\dot{\alpha}} \bar{\sigma}^{a \dot{\alpha} \alpha } \psi_{\alpha} 
-m^2 A \bar{A} - \frac{1}{2} m \psi^{\alpha} \psi_{\alpha} - \frac{1}{2} m \bar{\psi}_{\dot{\alpha}} \bar{\psi}^{\dot{\alpha}} 
\ee
where now we can see that supersymmetric masses have been raised.
The lesson from the above discussion is that, until integrating out the auxiliary sector, 
it is not obvious if there exists a mass term, 
and in a more general context, what is the form of the scalar potential.

Turning to supergravity, the above discussion is straightforwardly generalised and the same prosedure is followed.
The most general (two-derivative) superspace Lagrangian of chiral superfields
 coupled to supergravity is in superspace formalism
\footnote{Our framework  and  conventions are those of Wess and Bagger \cite{Wess:1992cp}.}
\be
\label{W&B}
{\cal{L}}_0=\frac{1}{\kappa^2} \int d^2 \Theta \, \,2 {\cal{E}} \left[ \frac{3}{8} \Big{(} \bar{{\cal{D}}}\bar{{\cal{D}}} -
 8 {\cal{R}} \Big{)}\,  e^{ - \frac{\kappa^2}{3} K(\Phi^i,\bar{\Phi}^{\bar{j}})} +\kappa^2 P(\Phi) \right] + h.c.
\ee
The hermitian function $K(\Phi^i,\bar{\Phi}^{\bar{j}})$ is the K\"ahler potential, 
$P(\Phi^i)$ is the superpotential (a holomorphic function of the chiral superfields $\Phi^i$) and $\kappa$ is 
proportional to the Planck length, which from now on will be set equal to 1. 
From the supergravity multiplet sector, $2{\cal{E}}$ is the usual chiral density employed to
 create supersymmetric Lagrangians, which  in the new $\Theta$ variables has the expansion
\be
\label{2Epsilon}
2{\cal{E}}=e \left\{ 1+ i\Theta \sigma^{a} \bar{\psi}_a -\Theta \Theta\Big{(} M^* +\bar{\psi}_a\bar{\sigma}^{ab}\bar{\psi}_b 
\Big{)} \right\}
\ee
in terms of  the vielbein ($e_m^a$), the gravitino ($\psi_m$) and 
 the complex scalar auxiliary field $M$. 
In addition, ${\cal{R}}$, the superspace curvature, is
 a chiral superfield which contains the Ricci scalar in its highest component. 
In the matter sector, $\Phi^i$ and $\bar{\Phi}^{\bar{j}}$ denote a set on chiral and anti-chiral superfields ($\bar{{\cal{D}}}_{\dot{\alpha}}\Phi^i=0 $, ${\cal{D}}_{\alpha}\bar{\Phi}^{\bar{j}}=0 $) whose components are defined via projection
\be
\label{chiral-def}
\nonumber
A^i&=&\Phi^i|_{\theta=\bar{\theta}=0}, \\
\chi^i_\alpha&=&\frac{1}{\sqrt{2}} {\cal{D}}_{\alpha} \Phi^i |_{\theta=\bar{\theta}=0}, \\
\nonumber
F^i&=&- \frac{1}{4}  {\cal{D}}{\cal{D}} \Phi^i|_{\theta=\bar{\theta}=0}.
\ee
After calculating the component form of (\ref{W&B}), integrating out the auxiliary fields and performing a Weyl rescaling of the gravitational field (accompanied by a redefinition of the fermionic fields), the pure bosonic Lagrangian reads
\be
\label{W&B-OnShell-rescaled}
e^{-1} {\cal{L}}_0=-\frac{1}{2} R - g_{i {\bar{j}}} \partial_a A^i \partial^a \bar{A}^{\bar{j}} - \ e^{K} \left[ g^{i {\bar{j}}} (D_iP) (D_{\bar{j}}\bar{P}) -3 P \bar{P}   \right].
\ee
Further details maybe found for example in \cite{Wess:1992cp}. Here 
\be
g_{i {\bar{j}}} = \frac{\partial^2 K(A,\bar{A})}{\partial A^i \partial\bar{A}^{\bar{r}}}
\ee
 is the positive definite K\"ahler metric, on the manifold parametrized by $A^i$ and $\bar{A}^{\bar{j}}$. 
Moreover, the K\"ahler space covariant derivatives are defined as follows
\be
\label{covariant-Khaler}
D_iP=P_i +K_iP
\ee
where in general we denote $f_i=\frac{\partial f }{ \partial A^i}$.
The Lagrangian (\ref{W&B-OnShell-rescaled}) is K\"ahler invariant as long as the superpotential  scales as 
\be
P(A^i)\rightarrow e^{-S(A^i)} P(A^i)
\ee
 under a K\"ahler transformation 
\be
K(A^i,\bar{A}^{\bar{j}})\rightarrow K(A^i,\bar{A}^{\bar{j}})+S(A^i)+\bar{S}(\bar{A}^{\bar{j}}).
\ee 
$S(A^i)$ and $\bar{S}(\bar{A}^{\bar{j}})$ are holomorphic functions of the complex coordinates.

Equaly important conventional methods for introducing scalar potentials 
is by { gauging the chiral models or by {\it D-terms}, 
the interested reader should consult \cite{Villadoro:2005yq}.

\section{F-Emergent Potential}

The idea of the emergent potentials is essentialy a generalization of the standard methods discused above.
The theory we are interested in, has a superspace Lagrangian of the form 
\be
{\cal{L}}={\cal{L}}_0+{\cal{L}}_{HD} \label{lang}
\ee
where ${\cal{L}}_0$ is the standard superspace supergravity Lagrangian given in eq.(\ref{W&B}) 
and \cite{Koehn:2012ar,Farakos:2012qu,Koehn:2012np}
\be
\label{KSB-Kahler-inv}
{\cal{L}}_{HD}=\int d^2 \Theta \, \, 2 {\cal{E}} \left\{ \frac{1}{8} 
\Big{( }\bar{{\cal{D}}}\bar{{\cal{D}}} - 8 {\cal{R}} \Big{)}   
\Lambda^{\bar{r}i\bar{n}j} 
\Big{[} 
\bar{{\cal{D}}}_{\dot{\alpha}} K_{i} 
{\cal{D}}_{\alpha}  K_{\bar{r}} 
\bar{{\cal{D}}}^{\dot{\alpha}}K_{j} 
{\cal{D}}^{\alpha}  K_{\bar{n}} 
\Big{]}  \right\} + h.c.
\ee
This Lagrangian was initially studied in global supersymmetry in \cite{Khoury:2010gb}.
It is important that ${\cal{L}}$ is manifestly both K\"ahler and (independently) super-Weyl invariant as has been shown in \cite{Farakos:2012qu}.
These two symmetry properties, although obviously they do not specify the form of the action, they are essential
in the consistency of the model as well as for the supergravity theory that it describes. 
As we will see, (\ref{KSB-Kahler-inv}) does not involve derivatives of the auxiliary fields, which are not propagating and can be integrated out. 
Equivalently, (\ref{KSB-Kahler-inv}) can be expressed in terms of the chiral superfields $\Phi^i$ as
\be
\label{KSB1}
{\cal{L}}_{HD}=\int d^2 \Theta \, \, 2 {\cal{E}} \left\{  \frac{1}{8} 
\Big{(} \bar{{\cal{D}}}\bar{{\cal{D}}} - 8 {\cal{R}} \Big{)}   \Lambda_{i\bar{r}j\bar{n}} \Big{[}
\bar{{\cal{D}}}_{\dot{\alpha}} {\bar{\Phi}}^{\bar{r}} {\cal{D}}_{\alpha} \Phi^i  \bar{{\cal{D}}}^{\dot{\alpha}}{\bar{\Phi}}^{\bar{n}} 
 {\cal{D}}^{\alpha} \Phi^j \Big{]}  \right\} + h.c.
\ee
where 
\be
K_{i\bar{r}}= \frac{\partial^2 K(\Phi,\bar{\Phi})}{\partial\Phi^i \partial\bar{\Phi}^{\bar{r}} }
\ee
 is the K\"ahler metric on the complex space spanned by the chiral and anti-chiral superfields and 
$\Lambda_{i\bar{r}j\bar{n}}$ represents a K\"ahler tensor. For example, one may choose
\be
\label{Lambda-ex}
{\Lambda}_{i\bar{r}j\bar{n}}= {\cal{G}}(\Phi,\bar{\Phi}) K_{i\bar{r}}K_{j\bar{n}}+ {\cal{H}}(\Phi,\bar{\Phi}) {\cal{R}}_{i\bar{r}j\bar{n}}
\ee
with ${\cal{G}}(\Phi,\bar{\Phi})$ and ${\cal{H}}(\Phi,\bar{\Phi})$ being some K\"ahler invariant hermitian functions and ${\cal{R}}_{i\bar{r}j\bar{n}}$ the K\"ahler space Riemann tensor defined as
\be
\label{Kahler-R}
{\cal{R}}_{i\bar{j}k\bar{l}}=\frac{\partial}{\partial {\Phi}^i}\frac{\partial}{\partial {\bar{\Phi}}^{\bar{j}}} K_{k\bar{l}} - K^{m\bar{n}} 
\left( \frac{\partial}{\partial {\bar{\Phi}}^{\bar{j}}} K_{m\bar{l}} \right) 
\left( \frac{\partial}{\partial {\Phi}^i} K_{k\bar{n}} \right).
\ee
The form (\ref{Lambda-ex}) implies some symmetries for the K\"ahler indices which, 
without loss of further generality, we will assume to be possessed by all the 
$\Lambda_{i\bar{r}j\bar{n}}$ to be considered in this work. 
Our next task is to extract the component 
field expression for the  Lagrangian (\ref{KSB1}), which 
after  superspace integration turns out to be
\be
\label{KSB-comp}
\nonumber
e^{-1}{\cal{L}}_{HD}=-16\ {\cal{U}}_{i\bar{r}j\bar{n}}\!\!
&\!\!&\!\!
\Big{(}F^{i} F^{j}\bar{F}^{\bar{r}}\bar{F}^{\bar{n}}
+\partial_a A^{i} \partial^a A^{j} \partial_b \bar{A}^{\bar{r}} \partial^b \bar{A}^{\bar{n}}
\\
&&
-F^{i} \bar{F}^{\bar{r}} \partial_a A^{j}\partial^a \bar{A}^{\bar{n}}
-F^{i} \bar{F}^{\bar{n}} \partial_a A^{j}\partial^a \bar{A}^{\bar{r}}\Big{)}.
\ee
for the pure bosonic sector. In (\ref{KSB-comp}) we have used the notation
\be
\label{LL}
{\cal{U}}_{i\bar{r}j\bar{n}}(A,\bar{A})=\Lambda_{i\bar{r}j\bar{n}}(\Phi,\bar{\Phi})\Big{|}_{\theta=\bar{\theta}=0}
\ee
Again it is easy to see that (\ref{KSB-comp}) is manifestly K\"ahler invariant.

In order to make the effect of the new  coupling (\ref{KSB-Kahler-inv})  more 
transparent we will consider now a theory with only \emph{ one chiral multiplet and no superpotential}. 
In this case, the  Lagrangian (\ref{lang}) 
is explicitly written as
\be
\label{FK1} 
{\cal{L}}=\int d^2 \Theta\,\, 2 {\cal{E}} \left\{ \Big{(}\bar{{\cal{D}}}\bar{{\cal{D}}} - 8 {\cal{R}}\Big{)} 
\left[\frac{3}{8} e^{-\frac{1}{3}K} +\frac{1}{8} \Lambda\ 
\bar{{\cal{D}}}_{\dot{\alpha}} {\bar{\Phi}} {\cal{D}}_{\alpha} \Phi \bar{{\cal{D}}}^{\dot{\alpha}}{\bar{\Phi}} {\cal{D}}^{\alpha} \Phi \right] \right\} + h.c.
\ee
with $\Lambda$ being an abbreviation for ${\Lambda}_{{\Phi}\bar{{\Phi}}{\Phi}\bar{{\Phi}}}$, a hermitian and 
K\"ahler invariant function of $\Phi$ and $\bar{\Phi}$. 
In component form, the bosonic sector of the Lagrangian (\ref{FK1}) turns out to be 
(after integrating out the auxiliary fields 
of the supergravity sector
and subsequently appropriately rescaling)
\be
\label{FK2} 
e^{-1}{\cal{L}}_{\rm{bos}}&=&
\nonumber -\frac{1}{2} R - g_{A {\bar{A}}} \partial_a A \partial^a \bar{A} + g_{A {\bar{A}}}\ e^{\frac{K}{3}} F\bar{F} \\
&& - 16\ {\cal{U}}\ \left\{
\phantom{\frac{X^X}{X^X}}\!\!\!\!\!\!\!\!\!\!\!
e^{\frac{2K}{3}} (F\bar{F})^2
+\partial_a A \partial^a A \partial_b \bar{A}\partial^b \bar{A}
- 2 e^{\frac{K}{3}} F\bar{F}\partial_a A\partial^a \bar{A} \,\right\}
\ee
where ${\cal{U}}$ is a hermitian K\"ahler invariant function of the scalar field (it is the lowest component of $\Lambda$, eq.(\ref{LL})).
 The equation of motion for $F$ is
\be
\label{Eq-F} 
\bar{F} \Big{(} g_{A {\bar{A}}}- 32\  {\cal{U}} e^{\frac{K}{3}} F \bar{F}
+ 32\  {\cal{U}} \partial_a A\partial^a \bar{A} \Big{)} =0
\ee
which can be easily solved for
\begin{enumerate}

 \item Standard solution:  \be \label{F0} F=0, \ee

  \item New solution:  \be  \label{Sol-F} 
F \bar{F} = e^{\frac{-K}{3}} \left( \frac{g_{A\bar{A}}}{32\ {\cal{U}}} +\partial_a A\partial^a \bar{A} \right).  \ee

\end{enumerate}
Here we should discuss the difference between the two solutions.
To make the point clear we first stress that the stability of the theory demands
\be
\label{K+}
g_{A\bar{A}} > 0
\\
\label{noghost}
{\cal{U}}   < 0.
\ee
Thus the standard solution (\ref{F0}) can always be realized, while the new solution  (\ref{Sol-F}) can only be realized in the presence of fluxes so that 
\be
\label{flux}
F \bar{F} = e^{\frac{-K}{3}} \left( \frac{g_{A\bar{A}}}{32\ {\cal{U}}} +\partial_a A\partial^a \bar{A} \right) > 0.
\ee
The on-shell Lagrangian for the conventional branch is
\be
\label{conv} 
e^{-1}{\cal{L}}_{\rm{bos}}= -\frac{1}{2} R - g_{A {\bar{A}}} \partial_a A \partial^a \bar{A} - 16\ {\cal{U}}\ \partial_a A \partial^a A \partial_b \bar{A}\partial^b \bar{A} 
\ee
where there is no scalar potential, as expected, since no superpotential was introduced.
The on-shell Lagrangian for the new branch is
\be
\label{FK3} 
e^{-1}{\cal{L}}_{\rm{bos}}= 
-\frac{1}{2} R 
+ \frac{ (g_{A {\bar{A}}})^2}{64\  {\cal{U}}} 
- 16\  {\cal{U}} \partial_a A \partial^a A \partial_b \bar{A}\partial^b \bar{A}
+ 16\  {\cal{U}} \partial_a A\partial^a \bar{A} \partial_b A\partial^b \bar{A}.
\ee
What has happened here has completely changed the dynamics of the theory. 
The minimal kinematic term for the scalar is lost and a scalar potential has \emph{emerged}
\be
\label{EP} 
{\cal{V}} = - \frac{1}{64} \frac{ (g_{A {\bar{A}}})^2}{  {\cal{U}}}.
\ee
From (\ref{noghost}) we see that the potential (\ref{EP}) is positive defined 
\be
{\cal{V}} > 0
\ee
and therefore the theory may only have de Sitter vacua. Another important property of the 
emerging potential is that it is not built from a holomorphic function. 
Moreover, the function \  ${\cal{U}}$ governs now the kinetic terms and in fact it was shown in \cite{Koehn:2012ar} that it  
 has to be negative to avoid tachionic states.
In the framework of new-minimal supergravity, consistent higher derivative terms which  
satsify the above restrictions have been considered 
\cite{ger-far}, 
but no scalar potential emerged in that case.

\section{Gauge Invariant F-Emergent Potential}

The Lagrangian (\ref{FK1}) can be straightforwardly be generalized to include gauge invariant 
interactions \cite{Bagger:1982ab}. In this case,  
the gauge invariant superspace  Lagrangian is
\be
\label{FKt-sups}
\nonumber
&&{\cal{L}}_{tot}=\int d^2 \Theta \,\,2 {\cal{E}} \left\{
\frac{3}{8} \Big{(} \bar{{\cal{D}}}\bar{{\cal{D}}} - 8 {\cal{R}} \Big{)} e^{ - {\tilde{K}}/3 } +\frac{1}{16g^2} {\cal{H}}_{(ab)}(\Phi) W^{(a)} W^{(b)} 
+ P(\Phi)\right. \nonumber  \\
&&\hspace{3cm}\left.+ \frac{1}{8} 
\Big{(} \bar{{\cal{D}}}\bar{{\cal{D}}} - 8 {\cal{R}} \Big{)} \left[ 
{\tilde{\Lambda}}^{\bar{r}i\bar{n}j}
\ 
\bar{{\cal{D}}}_{\dot{\alpha}} {\tilde{K}}_{i} 
{\cal{D}}_{\alpha}  {\tilde{K}}_{\bar{r}} 
\bar{{\cal{D}}}^{\dot{\alpha}} {\tilde{K}}_{j} 
{\cal{D}}^{\alpha}  {\tilde{K}}_{\bar{n}}  
\right] \right\}+ h.c.
\ee
where
\be
\label{Kahler-tilde}
{\tilde{K}}=K(\Phi,\bar{\Phi}) + \Gamma(\Phi,\bar{\Phi},V),
\ee
and
\be
\label{Gamma-WZgauge}
\Gamma(\Phi,\bar{\Phi},V)=V^{(a)}{\cal{D}}^{(a)} + \frac{1}{2} g_{i {\bar{r}}}  X^{i(a)} {\bar{X}}^{\bar{r}(b)} V^{(a)} V^{(b)}.
\ee
In addition, as usual, $V^{(a)}$ is the supersymmetric Yang-Mills vector multiplet and 
\be
W_{\alpha}=W_{\alpha}^{(a)}T^{(a)}=-\frac{1}{4}\Big{(} \bar{{\cal{D}}}\bar{{\cal{D}}} - 8 {\cal{R}} \Big{)} e^{-V}{\cal{D}}_{\alpha}e^{V}
\ee
is the gauge invariant chiral superfield containing the gauge field strength. The holomorphic function $ {\cal{H}}_{(ab)} $
is included for generality, but in what follows we will consider $ {\cal{H}}_{(ab)} = \delta_{(ab)} $. Expression (\ref{Gamma-WZgauge}) is calculated in the Wess-Zumino gauge, ${\cal{D}}^{(a)}$ 
are the so-called Killing potentials whereas  $X^{i(a)}$ and ${\bar{X}}^{\bar{r}(b)}$ are the components of the 
holomorphic Killing vectors that generate the isometries of the K\"ahler manifold. 
The Killing vectors and the Killing potential are connected via
\be
\label{X-D}
g_{i {\bar{r}}}{\bar{X}}^{\bar{r}(a)}=i\frac{\partial}{\partial a^i}{\cal{D}}^{(a)},\\
g_{i {\bar{r}}}X^{i(a)}=-i\frac{\partial}{\partial {\bar{a}}^{\bar{r}}}{\cal{D}}^{(a)}
\ee
where $a^i$ and ${\bar{a}}^{\bar{r}}$ are the K\"ahler space complex co-ordinates. We note that the ${\cal{D}}^{(a)}$ that correspond to some $U(1)$ gauged symmetry are only determined up to a constant 
$\xi$, which is the analog for the Fayet-Iliopoulos D-term in supergravity. Now ${\tilde{\Lambda}}^{\bar{r}i\bar{n}j} $ 
has to respect all the isometries of the K\"ahler manifold. 
Again, following the standard procedure,  the bosonic part of the Lagrangian (\ref{FKt-sups}) 
turns out to be
\be
\label{FKt-comp}
\nonumber
e^{-1}{\cal{L}}_{tot}= && -\frac{1}{2} R - g_{i {\bar{r}}} {\tilde{D}}_m A^i {\tilde{D}}^{m} \bar{A}^{\bar{r}}
 +e^{\frac{K}{3}} g_{i {\bar{r}}} F^i  \bar{F}^{\bar{r}} \\ 
\nonumber
&&  -\frac{1}{16g^2} F_{mn}^{(a)} F^{mn(a)}    -\frac{1}{2} g^2 \big{(} {\cal{D}}^{(a)} \big{)}^2        \\
&& - \ e^{\frac{2K}{3}} \Big{(} F^i D_i P + \bar{F}^{\bar{r}} D_{\bar{r}} \bar{P}\Big{)}   + 3 e^K P \bar{P} \\ 
\nonumber
&&-16\ \tilde{{\cal{U}}}_{i\bar{r}j\bar{n}} \Big{(}e^{\frac{2K}{3}}F^{i} F^{j}\bar{F}^{\bar{r}}\bar{F}^{\bar{n}}
+{\tilde{D}}_a A^{i} {\tilde{D}}^{a} A^{j} {\tilde{D}}_{b} \bar{A}^{\bar{r}} {\tilde{D}}^{b} \bar{A}^{\bar{n}} \\
\nonumber
&&\ \ \ \ \ \ \ \ \ \ -e^{\frac{K}{3}}F^{i} \bar{F}^{\bar{r}} {\tilde{D}}_a A^{j}{\tilde{D}}^{a} \bar{A}^{\bar{n}}
-e^{\frac{K}{3}}F^{i} \bar{F}^{\bar{n}} {\tilde{D}}_a A^{j}{\tilde{D}}^{a} \bar{A}^{\bar{r}}\Big{)}.
\ee
We note that  
\be
{\tilde{D}}_{c} A^{j} =\partial_c A^{j} - \frac{1}{2} B_{c}^{(a)} X^{j}_{(a)} 
\ee
is the covariant derivative and $B_{c}^{(a)}$ is a vector field (belonging to the $V^{(a)}$ 
vector multiplet) that corresponds to the gauged isometries, with 
 field strength $F_{mn}^{(a)}$.

In order to illustrate the properties of the {\it{emergent potential}} in the case of gauged models, 
our example will be a single chiral multiplet with \emph{no} superpotential. In this case the Lagrangian (\ref{FKt-comp}) is
\be
\label{FKt-comp-single}
\nonumber
e^{-1}{\cal{L}}_{tot}= && -\frac{1}{2} R - g_{A {\bar{A}}} {\tilde{D}}_m A {\tilde{D}}^{m} \bar{A}
 +e^{\frac{K}{3}} g_{A {\bar{A}}} F  \bar{F} \\ 
&&  -\frac{1}{16g^2} F_{mn}^{(a)} F^{mn(a)}    -\frac{1}{2} g^2 \big{(} {\cal{D}}^{(a)} \big{)}^2        \\
\nonumber
&&-16\ \tilde{{\cal{U}}} \Big{(}e^{\frac{2K}{3}} (F\bar{F})^2
+{\tilde{D}}_a A {\tilde{D}}^{a} A {\tilde{D}}_{b} \bar{A} {\tilde{D}}^{b} \bar{A} - 2\ e^{\frac{K}{3}}F 
\bar{F} {\tilde{D}}_a A{\tilde{D}}^{a} \bar{A} \Big{)}.
\ee
The single auxiliary field $F$ can now be eliminated from (\ref{FKt-comp-single}) by its equations of motion, leading to 
\be
\label{Sol-FG} 
F \bar{F} = e^{\frac{-K}{3}} \left( \frac{g_{A\bar{A}}}{32\  \tilde{{\cal{U}}}} +{\tilde{D}}_a A {\tilde{D}}^a \bar{A} \right).
\ee
Plugging (\ref{Sol-FG}) back in (\ref{FKt-comp-single}), we can easily read-off the potential for the gauged model
which turns out to be
\be
\label{UpliftedEP} 
{\cal{V}} = \frac{1}{2} g^2 \left({\cal{D}}^{(a)} \right)^2- \frac{(g_{A\bar{A}})^2}{64\  \tilde{{\cal{U}}}}
\ee
with $\tilde{{\cal{U}}}=\tilde{{\cal{U}}}_{A\bar{A}A\bar{A}}$, a K\"ahler-space 
tensor that respects all the isometries of the gauged group. 
For a first example we will take a flat model with K\"ahler potential 
\be
K=a \bar{a} +d\, 
\ee
which leads to
 \be
g_{a\bar{a}}=1\, , ~~~~{\cal{R}}_{a\bar{a}a\bar{a}}=0
\ee
The   $U(1)$ Killing potential is
\be
 D^{(1)}=a \bar{a} + \xi
\ee
where the parameter $\xi$ corresponds to the aforementioned freedom to 
shift the $U(1)$ Killing potential. When we promote $a$ and $\bar{a}$ to the superfields $\Phi$ and $\bar{\Phi}$,
 our K\"ahler potential $K$ together with the counter term $\Gamma$ become
\be
\label{Kahler-tilde-U(1)}
{\tilde{K}}_{U(1)}=\Phi \bar{\Phi} + V \Phi \bar{\Phi} +\frac{1}{2} V^2 \Phi \bar{\Phi} +d +V \xi.
\ee
The bosonic part of our Lagrangian in component form then turns out to be
\be
\label{FKt-comp-U(1)}
\nonumber
e^{-1}{\cal{L}}_{U(1)}= && -\frac{1}{2} R -\frac{1}{16g^2} F_{cd} F^{cd} \\
&&-16\ \tilde{{\cal{U}}} {\tilde{D}}_a A {\tilde{D}}^{a} A {\tilde{D}}_{b} \bar{A} {\tilde{D}}^{b} \bar{A} + 16\  \tilde{{\cal{U}}} {\tilde{D}}_a A {\tilde{D}}^{a} \bar{A} {\tilde{D}}_{b} A {\tilde{D}}^{b} \bar{A}  \\ \nonumber
&&-\frac{1}{2} g^2 \left( A \bar{A} + \xi  \right)^2 + \frac{1}{64\  \tilde{{\cal{U}}}},
\ee
with ${\tilde{D}}_{m} A =\partial_m A + \frac{i}{2} B_{m} A $. Then the  scalar potential is
\be
\label{EP-U(1)} 
{\cal{V}} = \frac{1}{2} g^2 \big{(} D^{(a)} \big{)}^2- \frac{1}{64 \, \tilde{{\cal{U}}}}.
\ee
A simple choice for $\tilde{{\cal{U}}}$ could  be 
\be
\tilde{{\cal{U}}}=mg_{A\bar{A}}g_{A\bar{A}}=m < 0 \, ,
\ee
 where  $m$ is a negative constant. 
It is again important to emphasise that $m$ now governs the kinematics of the scalar fields, 
and that the condition
\be
\label{FG-condition} 
F \bar{F} = e^{\frac{-K}{3}} \left( \frac{g_{A\bar{A}}}{32\  \tilde{{\cal{U}}}} +{\tilde{D}}_a A {\tilde{D}}^a \bar{A} \right) >0
\ee 
has to hold for the theory to be consistent.

\section{D-Emergent Potential}

Higher derivative interactions are not restricted only to scalar fields.
In fact we will show that an equivalent method as before can be followed which
again leads to a scalar potential.
Now the auxiliary fields that are integrated out are the ones of the 
vector multiplet, the ``$D$'' fields.

The higher derivative term we want to discuss is (in superspace) 
\be
\label{D-em}
{\cal{L}}_{gHD}=   \int d^2 \Theta \,\,2 {\cal{E}} 
\Big{(}\bar{{\cal{D}}}\bar{{\cal{D}}} - 8 {\cal{R}}\Big{)} 
(-\frac{1}{4} {\cal{J}}_{ab}(\Phi,\bar{\Phi}) W^{(a)} W^{(b)}   
{\cal{Y}}_{cd}(\Phi,\bar{\Phi}) \bar{W}^{(c)} \bar{W}^{(d)}  ) 
+ h.c. 
\ee
The superfields ${\cal{J}}_{ab}(\Phi,\bar{\Phi})$ and ${\cal{Y}}_{cd}(\Phi,\bar{\Phi})$ are functions of the various chiral superfields that are present in our theory,
the only restriction is that they should transform correctly under the gauge group.
The bosonic sector of Lagrangian (\ref{D-em}) after performing the superspace integration is
\be
\nonumber
e^{-1}{\cal{L}}_{gHD} &=& [J_{ab} \bar{Y}_{cd} + \bar{J}_{ab} Y_{cd} ]  \times 
\\
\nonumber
&\{& \frac{1}{4}  F^{dc(a)} F_{dc}^{(b)} F^{ab(c)} F_{ab}^{(d)} 
- \frac{1}{2}  F^{dc(a)} F_{dc}^{(b)} D^{(c)} D^{(d)} 
- \frac{1}{2}  D^{(a)} D^{(b)} F^{ab(c)} F_{ab}^{(d)} 
\\
\label{D-em-comp}
&& + D^{(a)} D^{(b)}  D^{(c)} D^{(d)} 
+ \frac{1}{16} \epsilon^{abcd} F_{ab}^{(a)} F_{cd}^{(b)} \epsilon^{efgh} F_{ef}^{(c)} F_{gh}^{(d)} \ \}.
\ee
Here $J_{ab}={\cal{J}}_{ab}|$ and $Y_{ab}={\cal{Y}}_{ab}|$.
Moreover for the gauge sector we will consider a more general coupling allowing for a kinetic gauge function as well.
The standard kinetic term for the gauge fields is
\be
\label{D-kin-gauge}
{\cal{L}}_{g0}=   \int d^2 \Theta \,\,2 {\cal{E}} 
{\cal{H}}_{(ab)}(\Phi) W^{(a)} W^{(b)} 
+ h.c. 
\ee
and the bosonic sector in components reads
\be
\label{D-kin-gauge-comp}
e^{-1} {\cal{L}}_{g0}=  [ H_{(ab)} + \bar{H}_{(ab)} ] 
\{ - \frac{1}{2} F^{dc(a)} F_{dc}^{(b)} 
- \frac{i}{4} \epsilon^{abcd} F_{ab}^{(a)} F_{cd}^{(b)} 
+ D^{(a)} D^{(b)} \}
\ee
with $H_{ab}={\cal{H}}_{ab}|$.
Up to now the most general Lagrangian in superspace reads
\be
\label{F+D-sups}
\nonumber
&&{\cal{L}}_{tot}=\int d^2 \Theta \,\,2 {\cal{E}} \left\{
\frac{3}{8} \Big{(} \bar{{\cal{D}}}\bar{{\cal{D}}} - 8 {\cal{R}} \Big{)} e^{ - {\tilde{K}}/3 } 
+ {\cal{H}}_{(ab)}(\Phi) W^{(a)} W^{(b)} 
+ P(\Phi)\right. \nonumber  \\
&&\hspace{3cm}\left.+ \frac{1}{8} 
\Big{(} \bar{{\cal{D}}}\bar{{\cal{D}}} - 8 {\cal{R}} \Big{)} \left[ 
{\tilde{\Lambda}}^{\bar{r}i\bar{n}j}
\ 
\bar{{\cal{D}}}_{\dot{\alpha}} {\tilde{K}}_{i} 
{\cal{D}}_{\alpha}  {\tilde{K}}_{\bar{r}} 
\bar{{\cal{D}}}^{\dot{\alpha}} {\tilde{K}}_{j} 
{\cal{D}}^{\alpha}  {\tilde{K}}_{\bar{n}}  
\right] \right.
\\
\nonumber
&&\left.\hspace{3cm}-\frac{1}{4}\Big{(}\bar{{\cal{D}}}\bar{{\cal{D}}} - 8 {\cal{R}}\Big{)} 
[ {\cal{J}}_{ab}(\Phi,\bar{\Phi}) W^{(a)} W^{(b)}   
{\cal{Y}}_{cd}(\Phi,\bar{\Phi}) \bar{W}^{(c)} \bar{W}^{(d)}  ] 
 \right\}+ h.c.
\ee

Finally, in order to study the properties of this new term, 
let us consider a very simple example of 
a single $U(1)$ group 
and a single uncharged (under this $U(1)$) chiral multiplet.
The higher derivative terms will be only for the gauge sector.
Our Lagrangian, in component form reads
\be
\nonumber
e^{-1}{\cal{L}}_{ex}= && -\frac{1}{2} R 
- g_{A {\bar{A}}} \partial_m A \partial^m \bar{A} 
 +[ H(A) + \bar{H}(\bar{A}) ] 
\{ - \frac{1}{2} F^{dc} F_{dc}
- \frac{i}{4} \epsilon^{abcd} F_{ab} F_{cd} 
+ D^2 \} 
\\
\label{U1}
&& +[J \bar{Y} + Y \bar{J} ] 
\{ \frac{1}{4} (F^{dc} F_{dc})^2 - F^{dc} F_{dc} D^2 + \frac{1}{16} (\epsilon^{abcd} F_{ab} F_{cd} )^2 +D^4 \}.
\ee
Here $J$ and $Y$ are positive definite gauge invariant functions of $A$ and $\bar{A}$.
Now we can easily solve the auxiliary $D$ equations of motion to find two solutions
\begin{enumerate}

 \item Standard solution:  \be D=0, \ee

  \item New solution:  \be D^2 = \frac{1}{2} F^{dc} F_{dc} 
  -\frac{1}{2} \frac{H + \bar{H}}{ J \bar{Y} + Y \bar{J}}. \ee

\end{enumerate}
The first one is the standard supersymmetric solution and has been also studied in \cite{Cecotti:1986jy} in the presence of higher derivatives. 
The new solution can only be consistently realized in the presence of magnetic fluxes so that
\be 
D^2 = \frac{1}{2} F^{dc} F_{dc} 
  -\frac{1}{2} \frac{H + \bar{H}}{ J \bar{Y} + Y \bar{J}} > 0. 
\ee
Eventually the on-shell theory will be
\be
\nonumber
e^{-1}{\cal{L}}_{ex}= && -\frac{1}{2} R 
- g_{A {\bar{A}}} \partial_m A \partial^m \bar{A} 
-\frac{1}{4} \frac{(H + \bar{H})^2}{J \bar{Y} + Y \bar{J}}
- \frac{i}{4} [ H(A) + \bar{H}(\bar{A}) ] \epsilon^{abcd} F_{ab} F_{cd} 
\\
\nonumber
&& + \frac{1}{16} [J \bar{Y} + Y \bar{J} ] 
 (\epsilon^{abcd} F_{ab} F_{cd} )^2  
\\
\nonumber
= && -\frac{1}{2} R 
- g_{A {\bar{A}}} \partial_m A \partial^m \bar{A} 
-\frac{1}{4} \frac{(H + \bar{H})^2}{J \bar{Y} + Y \bar{J}}
- \frac{i}{4} [ H(A) + \bar{H}(\bar{A}) ] \epsilon^{abcd} F_{ab} F_{cd} 
\\
\label{U1-onshell}
&& +[J \bar{Y} + Y \bar{J} ] 
\{ - \frac{1}{2} (F^{dc} F_{dc})^2  +   F_{ab}  F^{bc} F_{cd} F^{da}  \} .
\ee
It is easy to see that there is a positive definite emergent potential 
due to integrating out of the $D$ auxiliary field
\be
\label{D-Em}
{\cal V}(A,\bar{A}) = \frac{1}{4} \frac{(H + \bar{H})^2}{J \bar{Y} + Y \bar{J}}.
\ee
A simple example can be given by a gauge kinetic function 
\be
H= A^2
\ee
with $a$, $b$ two real positive constants
\be
\nonumber
J= a >0, \ \ \ Y= b >0.
\ee
The potential will be
\be
\label{D-Am}
{\cal V}(A,\bar{A}) = \frac{(A^2 + \bar{A}^2)^2}{8ab}.
\ee
This novel feature of gauge fields higher derivatives has not been studied before and deserves further investigation.

Summarizing, the well-known standard form of the N = 1 scalar potential is
restricted to the
two-derivative  level. Higher derivative interaction modify its form. In
fact,
when  higher-derivatives are introduced, an
emerging scalar potential appears even if there is no superpotential to
start with. There are two types of emerging potential, F- and D-type.
F-emerging potentials result by integrating out auxiliaries of chiral
multiplets whereas, D-emerging potentials come from the integration of
auxiliaries in vector multiplets. As a general rule,  emerging potentials
are positive defined with de Sitter ground state, indicating supersymmetry
breaking.

\end{document}